\documentclass[fleqn,twoside]{article}
\usepackage{espcrc2}

\usepackage{graphicx}

\title{Theory of Scalars}

\author{N.N. Achasov \address[MCSD]{Laboratory of Theoretical Physics,
S.L. Sobolev Institute for Mathematics, Academician Koptiug
Prospekt, 4, Novosibirsk, 630090, Russia}, A.V. Kiselev
\addressmark{\tt}, and G.N. Shestakov
\addressmark{\tt}}

\begin{document}

\begin{abstract}
Outline: 1. Introduction, 2. Confinement, chiral dynamics and light
scalar mesons, 3. Chiral shielding of the $\sigma(600)$, chiral
constraints (the CGL band), the $\sigma(600)$ and the $f_0(980)$ in
$\pi\pi\to\pi\pi$, $\pi\pi\to K\bar K$, $\phi\to\gamma\pi^0\pi^0$,
4. The $\phi$ meson radiative decays on light scalar resonances, 5.
Light scalars in $\gamma\gamma$ collisions.

Evidence for four-quark components of light scalars is given. The
priority of Quantum Field Theory in revealing the light scalar
mystery is emphasized.
\end{abstract}

\maketitle

\section{Introduction}
The scalar channels in the region up to 1 GeV became a stumbling
block of QCD. The point is that both perturbation theory and sum
rules do not work in these channels because there are not solitary
resonances in this region.

At the same time the question on the nature of the light scalar
mesons is major for  understanding the mechanism of the chiral
symmetry realization, arising from the confinement, and hence for
understanding the confinement itself.

\section{Place in QCD, chiral limit, confinement, and \boldmath $\sigma$ models}
The QCD Lagrangian is given by \\ $L$=$-\frac{1}{2}Tr\left
(G_{\mu\nu}(x)G^{\mu\nu}(x)\right)+\bar q(x)(i\hat{D}-M)q(x),$\\
$M$ is a diagonal matrix of quark masses, $\hat{D}$=$\gamma^\mu
D_\mu, D_\mu$=$\partial_\mu+ig_0G_\mu (x).$ $M$ mixes left and right
spaces. But in chiral limit, $M_{ff}$\,$\to$\,0, these spaces
separate realize $U_L(3)\times U_R(3)$ flavour symmetry, which,
however, is broken by the gluonic anomaly up to
$U_{\mbox{\scriptsize{vec}}}(1)\times SU_L(3)\times SU_R(3).$ As
experiment suggests, confinement forms colourless observable
hadronic fields and spontaneous breaking of chiral symmetry with
massless pseudoscalar fields. There are two possible scenarios for
QCD at low energy. 1. Non-linear $\sigma$ model. 2. Linear $\sigma$
model (LSM). The experimental nonet of the light scalar mesons,
$f_0(600)$ (or $\sigma (600)$), $\kappa$(700-900), $a_0(980)$ and
$f_0(980)$ mesons, suggests the $U_L(3)\times U_R(3)$ LSM.

\section{History}
Hunting the light  $\sigma$ and $\kappa$ mesons had begun in the
sixties already and a preliminary information on the light scalar
mesons in 
PDG Reviews had appeared at that time. But long-standing
unsuccessful attempts to prove their existence in a conclusive way
entailed general disappointment and an information on these states
disappeared from PDG Reviews. One of principal reasons against the
$\sigma$ and $\kappa$ mesons was the fact that both $\pi\pi$ and
$\pi K$ scattering phase shifts do not pass over $90^0$ at putative
resonance masses.

\section{\boldmath $SU_L(2)\times SU_R(2)$ LSM,
$\pi\pi\to\pi\pi$ \cite{GML,AS94,AS07}} Situation changes when we
showed that in LSM there is a negative background phase which hides
the $\sigma$ meson in $\pi\pi\to\pi\pi$. It has been made clear that
shielding wide lightest scalar mesons in chiral dynamics is very
natural. This idea was picked up and triggered new wave of
theoretical and experimental searches for the $\sigma$ and $\kappa$
mesons. Our approximation is as follows (see Fig.\,1):

$T_0^{0(tree)}$ = $\frac{m_\pi^2-m_\sigma^2}{32\pi f^2_\pi}\left[
5-3\frac{m_\sigma^2-m_\pi^2}{m_\sigma^2-s}-2\frac{m_\sigma^2-
m_\pi^2}{s-4m_\pi^2}\right.$ \\ $\left. \times\ln\left
(1+\frac{s-4m^2_\pi}{m_\sigma^2}\right )\right],\
T^0_0=T_0^{0(tree)}$ $/[1-i\rho_{\pi\pi}\\ \times T_0^{0(tree)}]=
[e^{2i\left(\delta_{bg}+\delta_{res}\right)}$--$\,1]/(2i\rho_{\pi\pi})
$\,=\,$(e^{2i\delta^0_0}$--$\,1)/(2i\rho_{\pi\pi})$ = $T_{bg}+
e^{2i\delta_{bg}}T_{res},$ $T_{res}$ =
$[\sqrt{s}\Gamma_{res}(s)/\rho_{\pi\pi}]/[M^2_{res}-s+
\mbox{Re}\Pi_{res}(M^2_{res})-\Pi_{res}(s)]
=(e^{2i\delta_{res}}-1)/(2i\rho_{\pi\pi})\,,\
T_{bg}=(e^{2i\delta_{bg}}-1)/(2i\rho_{\pi\pi})
=\lambda(s)/[1-i\rho_{\pi\pi}\lambda(s)]$, $\lambda(s)
=\frac{m_\pi^2-m_\sigma^2}{32\pi f^2_\pi}\\
\times\left[5-2\frac{m_\sigma^2-m_\pi^2}{s-4m_\pi^2}\ln\left
(1+\frac{s-4m^2_\pi}{m_\sigma^2}\right)\right]$, $f_\pi$=92.4\,MeV,\\
$\mbox{Re}\Pi_{res}(s)$=
$-g_{res}^2(s)\lambda(s)\rho_{\pi\pi}^2/(16\pi)$, $\mbox{Im}
\Pi_{res}(s)$= $g_{res}^2(s)\rho_{\pi\pi}/(16\pi)$,
$g_{res}(s)$=$g_{\sigma\pi\pi}/|1-i\rho_{\pi\pi}\lambda(s)|$,
$M^2_{res}$=$m_\sigma^2 - \mbox{Re}\Pi_{res}(M^2_{res})$,
$\rho_{\pi\pi}$=$(1-4m_\pi^2/s)^{1/2}$,
$g_{\sigma\pi\pi}$=$(3/2)^{1/2}g_{\sigma\pi^+\pi^-}
$=$(3/2)^{1/2}(m^2_\pi-m^2_\sigma)/f_\pi$.
\begin{figure}\includegraphics[width=17pc,height=7pc]{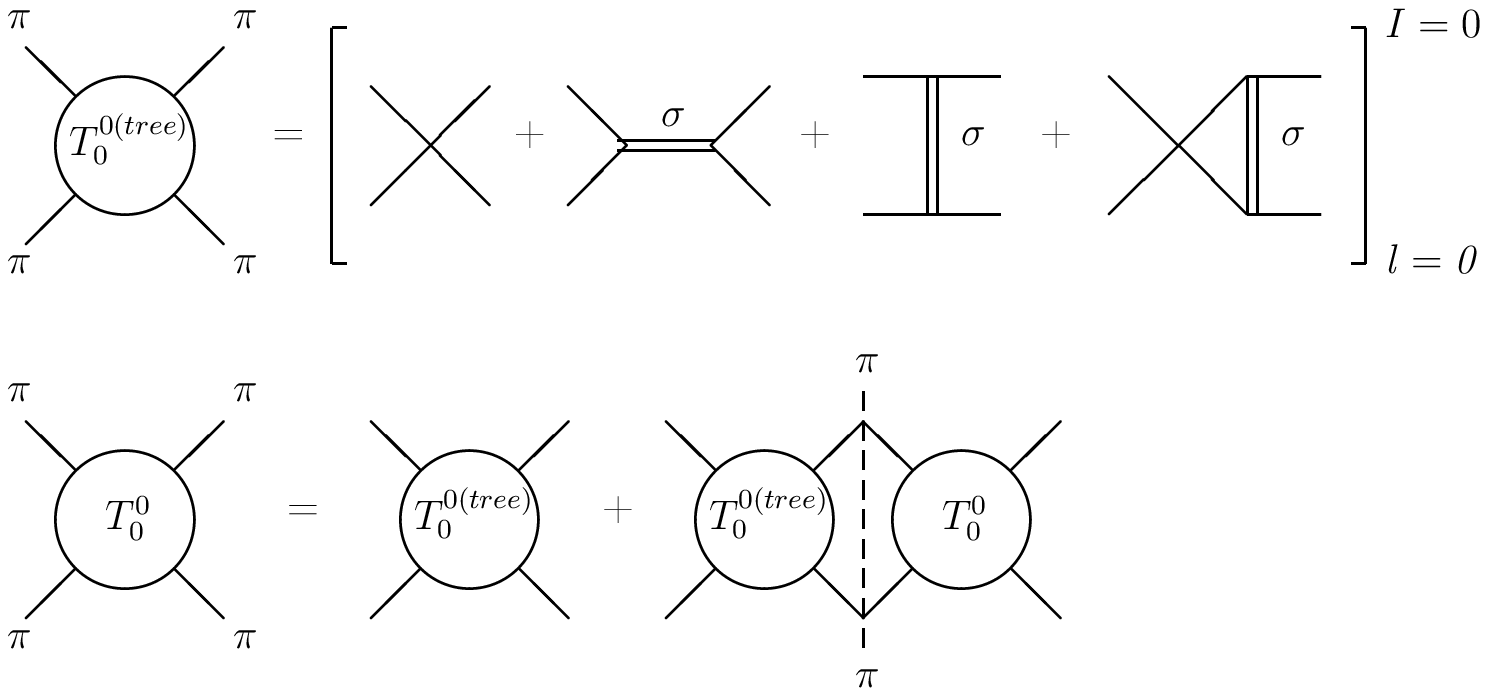}
\\ {\footnotesize Figure 1. The graphical representation of the $S$ wave $I=0$
$\pi\pi$ scattering amplitude $T^0_0$.}\end{figure}
$T^{2(tree)}_0=\frac{m_\pi^2-m_\sigma^2}{16\pi f^2_\pi}\left
[1-\frac{m_\sigma^2-m_\pi^2}{s-4m_\pi^2}\ln\left
(1+\frac{s-4m^2_\pi}{m_\sigma^2}\right)\right]$,\\
$\mbox{\qquad}T^2_0$=$T_0^{2(tree)}/[1-i\rho_{\pi\pi}T_0^{2(tree)}]$=$
\frac{e^{2i\delta_0^2}-1}{2i\rho_{\pi\pi}}$.

\section{Results in our approximation \cite{AS07}.}
$M_{res}$=$0.43$ GeV, $\Gamma_{res}(M^2_{res})$=$0.67$ GeV, $
m_\sigma$=$0.93$ GeV, $\Gamma^{renorm}_{res}(M^2_{res})
$=$\Gamma_{res}(M^2_{res})/(1+
d\mbox{Re}\Pi_{res}(s)/ds|_{s=M^2_{res}})=0.53\,\mbox{GeV}\,,$
$\Gamma_{res}(s)=\frac{g_{res}^2(s)}{16\pi\sqrt{s}}\rho_{\pi\pi}\,,$
$g_{res}(M^2_{res})/g_{\sigma\pi\pi}$=0.33, $a^0_0$=$0.18\,
m_\pi^{-1}$, $a^2_0$=$-0.04\, m_\pi^{-1}$, $(s_A)^0_0$=$0.45\,
m^2_\pi$, $(s_A)^2_0$=$2.02\, m^2_\pi\,.$

\section{Chiral shielding in \boldmath $\pi\pi\to\pi\pi$ \cite{AS07}}
\begin{figure}\includegraphics[width=17pc,height=7pc]{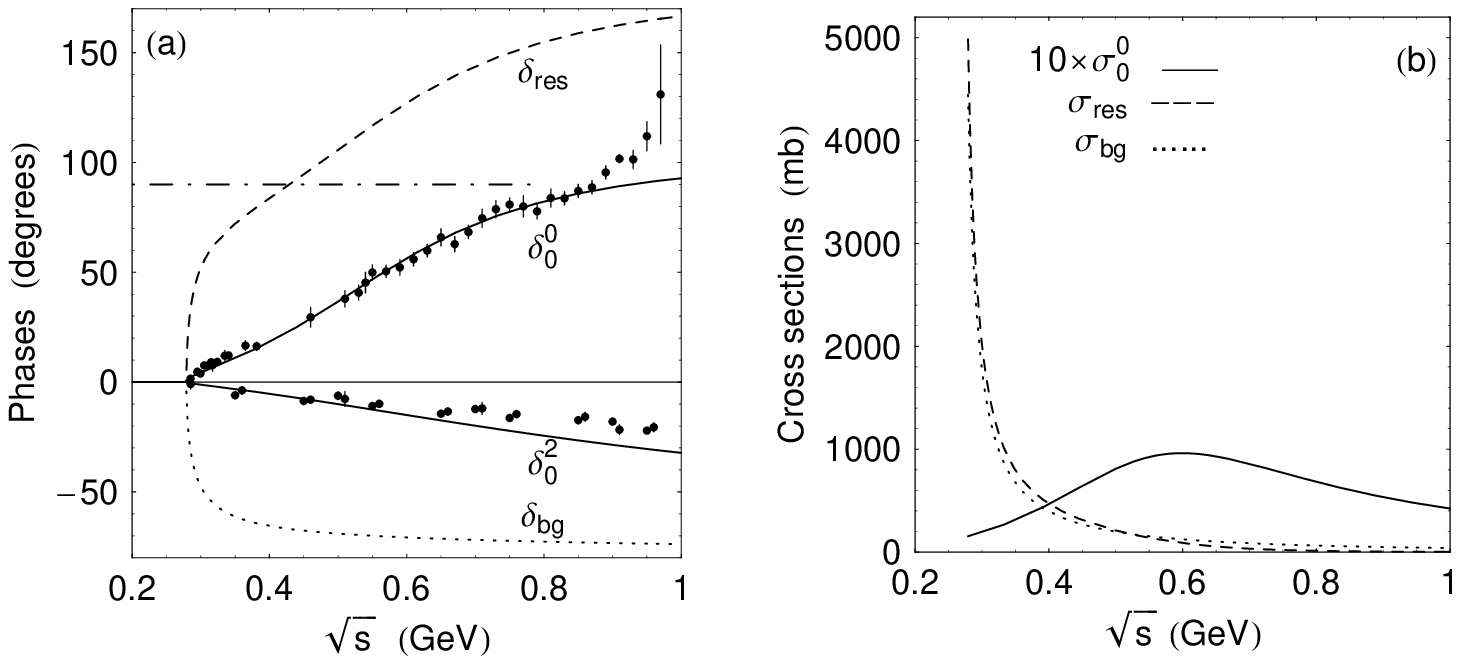}
\\ {\footnotesize Figure 2. The $\sigma$ model. Our approximation.
$\delta^0_0=\delta_{res}+\delta_{bg}$.
($\sigma^0_0,\sigma_{res},\sigma_{bg}$)=$32\pi(|T^0_0|^2,
|T_{res}|^2,|T_{bg}|^2)/s$.}\end{figure}

The chiral shielding of the $\sigma(600)$ meson is illustrated in
Fig. 2(a) with the help of the $\pi\pi$ phase shifts $\delta_{res}$,
$\delta_{bg}$, and $\delta^0_0$, and in Fig. 2(b) with the help of
the corresponding cross sections.

\section{The \boldmath $\sigma$ pole in $\pi\pi\to\pi\pi$ \cite{AS07}}
$T^0_0\to g^2_\pi/(s-s_R)\,,$ $ g^2_\pi$=$(0.12+i0.21)$ GeV$^2$,

$\sqrt{s_R}=M_R-i\Gamma_R/2=(0.52-i0.25)\ \mbox{GeV}.$\\
Considering the residue of the $\sigma$ pole in $T^0_0$ as the
square of its coupling constant to the $\pi\pi$ channel is not a
clear guide to understand the $\sigma$ meson nature for its great
obscure imaginary part.

\section{The \boldmath{$\sigma$} propagator \cite{AS07}}
$1/D_\sigma(s)$=$1/[M^2_{res}$--$s$+$\mbox{Re}\Pi_{res}(M^2_{res})$--$
\Pi_{res}(s)]$.\\  The $\sigma$ meson self-energy $\Pi_{res}(s)$ is
caused by the intermediate $\pi\pi$ states, that is, by the
four-quark intermediate states if we keep in mind that the
$SU_L(2)\times SU_R(2)$ LSM could be the low energy realization of
the two-flavour QCD. This contribution shifts the Breit-Wigner (BW)
mass greatly $m_\sigma- M_{res}$=\,0.50 GeV. So, half the BW mass is
determined by the four-quark contribution at least. The imaginary
part dominates the propagator modulus in the region
300\,MeV\,$<\sqrt{s}<$\,600\, MeV. So, the $\sigma$ field is
described by  its four-quark component at least in this energy
region.

\section{Chiral shielding in \boldmath $\gamma\gamma\to\pi\pi$ \cite{AS07}}
$\mbox{\ \ }T_S(\gamma\gamma\to\pi^+\pi^-)=
T_S^{Born}(\gamma\gamma\to\pi^+\pi^-)$\\ $\mbox{\ \ \ \ \ \ \ \
}+8\alpha I_{\pi^+\pi^-}T_S(\pi^+\pi^-\to\pi^+\pi^-)$\\ $=
T_S^{Born}(\gamma\gamma\to\pi^+\pi^-)+\frac{8\alpha}{3}
I_{\pi^+\pi^-}\left(2T_0^0 +T_0^2\right)$,\\
$T_S(\gamma\gamma\to\pi^0\pi^0)=8\alpha
I_{\pi^+\pi^-}\,T_S(\pi^+\pi^-\to\pi^0\pi^0)$\\ $\mbox{\ \ \ \ \ \ \
\ \ \ \ \ }=16\alpha I_{\pi^+\pi^-}(T_0^0 - T_0^2)/3$,\\
$I_{\pi^+\pi^-}= \frac{m^2_\pi}{s}
(\pi+i\ln\frac{1+\rho_{\pi\pi}}{1-\rho_{\pi\pi}})^2-1,\ \ s\geq
4m_\pi^2,$

$ T_S^{Born}(\gamma\gamma\to\pi^+\pi^-)=
(8\alpha/\rho_{\pi^+\pi^-})\mbox{Im}I_{\pi^+\pi^-}.$

Our results are shown in Fig. 3.

$\Gamma(\sigma$\,$\to$\,$\pi^+\pi^-$\,$\to$\,$\gamma\gamma,s)$=$
\frac{1}{16\pi\sqrt{s}}|g(\sigma$\,$\to$\,$\pi^+\pi^-$\,$\to $\\
$\gamma\gamma,s)|^2$, where
$g(\sigma$\,$\to$\,$\pi^+\pi^-$\,$\to$\,$\gamma\gamma,\,s)$=$
(\alpha/2\pi)$ $\times I_{\pi^+\pi^-}g_{res\,\pi^+\pi^-}(s)$; see
Fig. 4. So, the the $\sigma\to\gamma\gamma$ decay is described by
the triangle $\pi^+\pi^-$ loop diagram
$res\to\pi^+\pi^-\to\gamma\gamma$. Consequently, it is due to the
four-quark transition because we imply a low energy realization of
the two-flavour QCD by means of the the $SU_L(2)\times SU_R(2)$ LSM.
As Fig. 4 suggests, the real intermediate $\pi^+\pi^-$ state
dominates in $g(res\to\pi^+\pi^-\to\gamma\gamma)$ in the $\sigma$
region $\sqrt{s}<$\,0.6\,GeV. Thus the picture in the physical
region is clear and informative. But, what  about the pole in the
complex $s$ plane? Does the pole residue reveal the $\sigma$ indeed?

\section{The \boldmath $\sigma$  pole in $\gamma\gamma\to\pi\pi$ \cite{AS07}}
$\frac{1}{16\pi}\sqrt{\frac{3}{2}}\,T_S(\gamma\gamma\to\pi^0\pi^0)\to
g_\gamma g_\pi/(s-s_R)$,\\
$g_\gamma g_\pi$=$(-0.45 - i0.19)\cdot 10^{-3}\,\mbox{GeV}^2$,
$g_\gamma/g_\pi= (-1.61+i1.21)\cdot10^{-3}$,
$\Gamma(\sigma\to\gamma\gamma)$=$\frac{|g_\gamma |^2}{M_R}\approx
2\,\mbox{keV}.$ It is hard to believe that anybody could learn the
complex but physically clear dynamics of the $\sigma\to\gamma\gamma$
decay from the residues of the $\sigma$ pole.

\begin{figure}
\begin{center}
\begin{tabular}{c}
{\includegraphics[width=17pc,height=7pc]{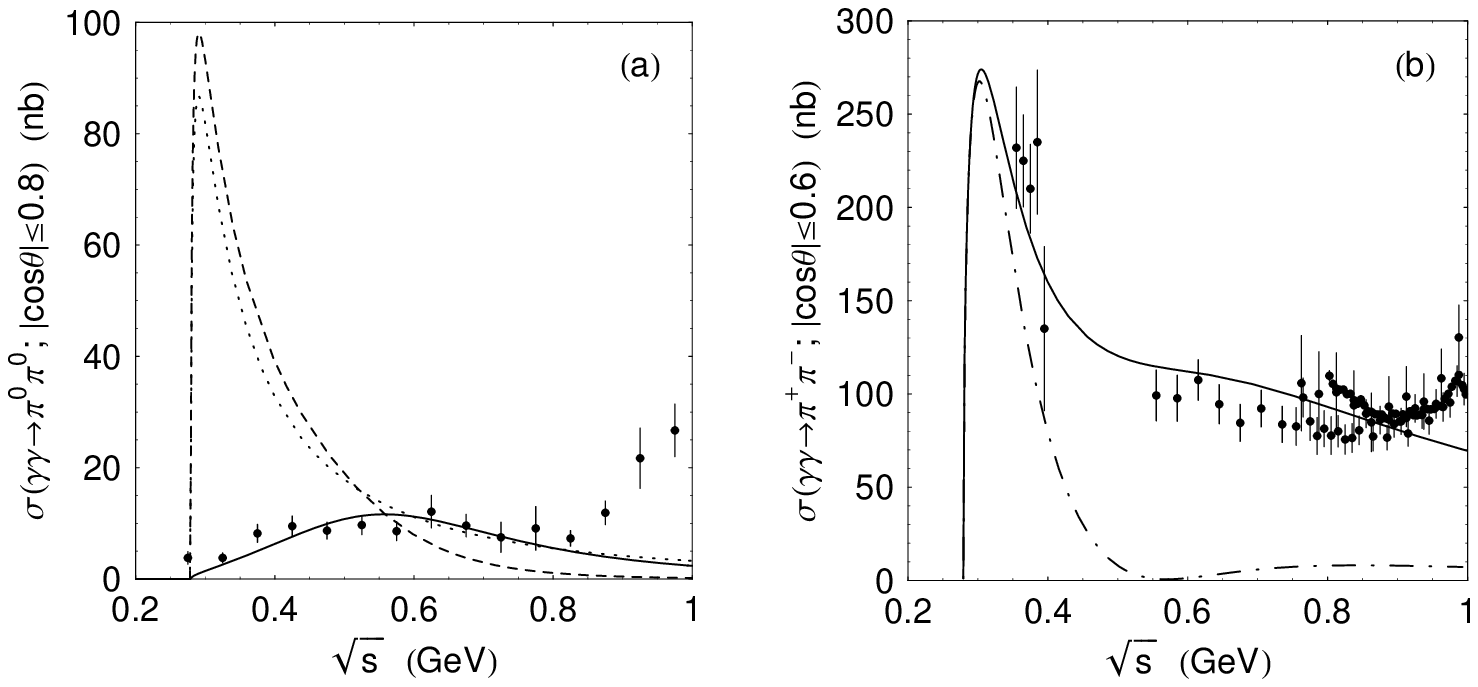}}
\end{tabular}\\ \end{center}
{\footnotesize Figure 3. (a) The solid, dashed, and dotted lines are
$\sigma_S(\gamma\gamma\to\pi^0\pi^0)$,
$\sigma_{res}(\gamma\gamma\to\pi^0\pi^0)$, and
$\sigma_{bg}(\gamma\gamma\to\pi^0\pi^0)$. (b) The dashed-dotted line
is $\sigma_S(\gamma\gamma\to\pi^+\pi^-)$, the solid line
includes the higher waves from $T^{Born}(\gamma\gamma\to\pi^+\pi^-)$.}\\
\begin{center}\begin{tabular}{c}
{\includegraphics[width=0.3\textwidth,height=6.7pc]{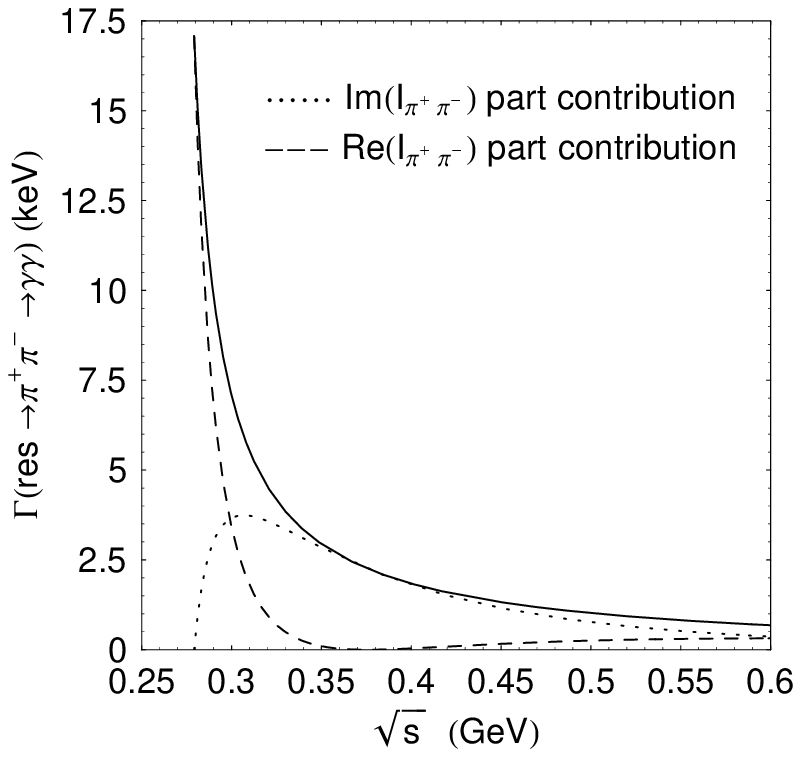}}
\end{tabular}\\ \end{center}
{\footnotesize Figure 4. The energy dependent width of the
$\sigma\to\pi^+\pi^- \to\gamma\gamma$ decay.}\end{figure}

\section{Discussion \cite{AS07,CCL,Ja,ADS1}}
Leutwyler and collaborators obtained $\sqrt{s_R}= M_R-i\Gamma_R/2
=\left(441^{+16}_{-8}-i272^{+12.5}_{-9}\right )\,\mbox{MeV}$ with
the help of the Roy equation. Our result agrees with the above only
qualitatively. $\sqrt{s_R}= M_R-i\Gamma_R/2
=(518-i250)\,\mbox{MeV}$. This is natural, because our approximation
gives only a semiquantitative description of the data at $\sqrt{s}<
0.4$ GeV. We do not regard also for   effects of the $K\bar K$
channel, the $f_0(980)$ meson, and so on, that is, do not consider
the $SU_L(3)\times SU_R(3)$ LSM.

Could the above scenario incorporates the primary lightest scalar
Jaffe four-quark  state? Certainly the direct coupling of this state
to $\gamma\gamma$ via neutral vector pairs ($\rho^0\rho^0$ and
$\omega\omega$), contained in its wave function, is negligible,
$\Gamma(q^2\bar q^2\to\rho^0\rho^0+\omega\omega\to\gamma\gamma)
\approx 10^{-3}$ keV, as we showed in 1982 \cite{ADS1}. But its
coupling to $\pi\pi$ is strong and leads to $\Gamma(q^2\bar
q^2\to\pi^+\pi^-\to\gamma\gamma)$ similar to
$\Gamma(res\to\pi^+\pi^-\to\gamma\gamma)$ in the above Fig. 4.

Let us add to $T_S(\gamma\gamma\to\pi^0\pi^0)$ the amplitude for the
the direct coupling of $\sigma$ to $\gamma\gamma$ conserving
unitarity $T_{direct}(\gamma\gamma\to\pi^0\pi^0)=sg^{(0)}_{\sigma
\gamma\gamma}g_{res}(s)e^{i\delta_{bg}}/D_{res}(s)$, where
$g^{(0)}_{\sigma\gamma\gamma}$ is the direct coupling constant of
$\sigma$ to $\gamma\gamma$, the factor $s$ is caused by gauge
invariance. Fitting the $\gamma\gamma\to\pi^0\pi^0$ data gives a
negligible value of $g^{(0)}_{\sigma\gamma\gamma}$,\\
$\Gamma^{(0)}_{\sigma\gamma\gamma}=
|M^2_{res}g^{(0)}_{\sigma\gamma\gamma}|^2/(16\pi M_{res})\approx
0.0034$ keV, in astonishing agreement with our prediction
\cite{ADS1}.

The majority of current investigations of the mass spectra in scalar
channels do not study particle production mechanisms. Because of
this, such investigations are essentially preprocessing experiments,
and  the derivable information is very relative. Nevertheless, the
progress in understanding the particle production mechanisms could
essentially help us reveal the light scalar meson nature.

\section{Troubles and expectancies}
In theory the principal problem is impossibility to use the linear
$\sigma$ model in the tree level approximation inserting widths into
$\sigma$ meson propagators because such an approach breaks the both
unitarity and Adler self-consistency conditions. Strictly speaking,
the comparison with the experiment requires the non-perturbative
calculation of the process amplitudes. Nevertheless, now there are
the possibilities to estimate odds of the $U_L(3)\times U_R(3)$ LSM
to underlie physics of light scalar mesons in phenomenology. Really,
even now there is a body of information about the $S$ waves of
different two-particle pseudoscalar states. As for theory, we know
quite a lot about the scenario under discussion: the nine scalar
mesons, the putative chiral shielding of the $\sigma(600)$ and
$\kappa$(700-900) mesons, the unitarity, analyticity and Adler
self-consistency conditions. In addition, there is the light scalar
meson treatment motivated by field theory.

\section{Phenomenological chiral shielding \cite{AK06}}
$g_{\sigma\pi^+\pi^-}^2/4\pi$=0.99 GeV$^2$,$\ g_{\sigma
K^+K^-}^2/4\pi$=2$\cdot10^{-4}$ GeV$^2$,
$g_{f_0\pi^+\pi^-}^2/4\pi$=0.12 GeV$^2$, $g_{f_0
K^+K^-}^2/4\pi$=1.04 GeV$^2$. The BW masses and width: $m_{f_0}$=989
MeV, $m_\sigma$=679 MeV, $\Gamma_\sigma$=498 MeV. The $l$=$I$=0
$\pi\pi$ scattering length $a^0_0$ =$0.223\, m^{-1}_{\pi^+}$. Figure
5 illustrates the excellent agreement our phenomenological treatment
with the experimental and theoretical data.
\begin{figure}
\begin{center}
\begin{tabular}{cc}
\includegraphics[width=6pc]{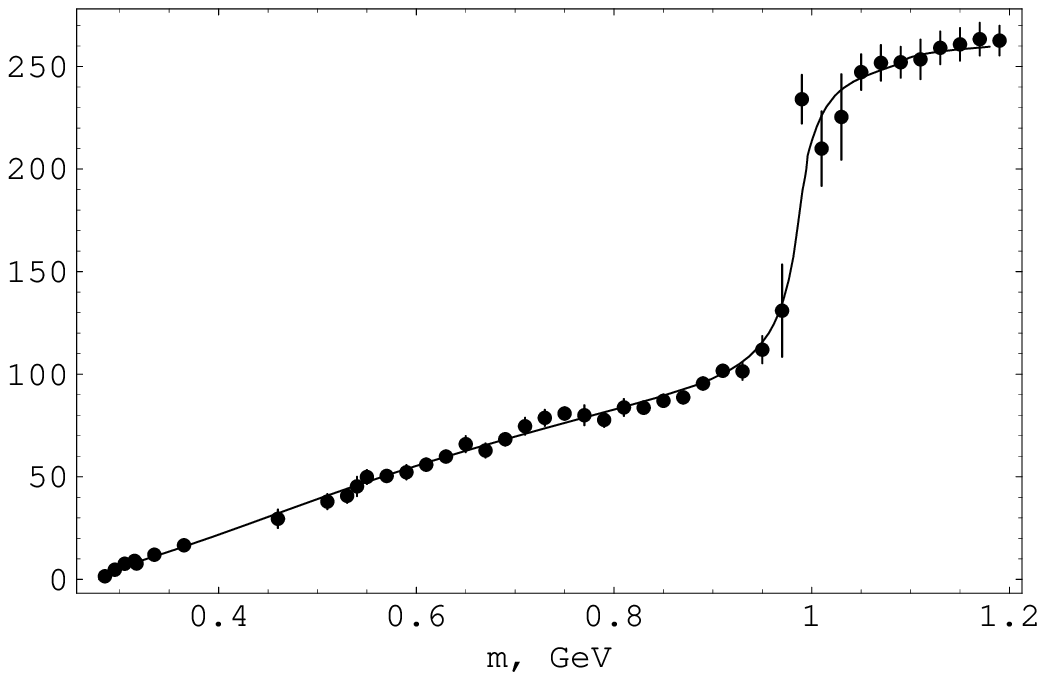}\ \
\includegraphics[width=6pc]{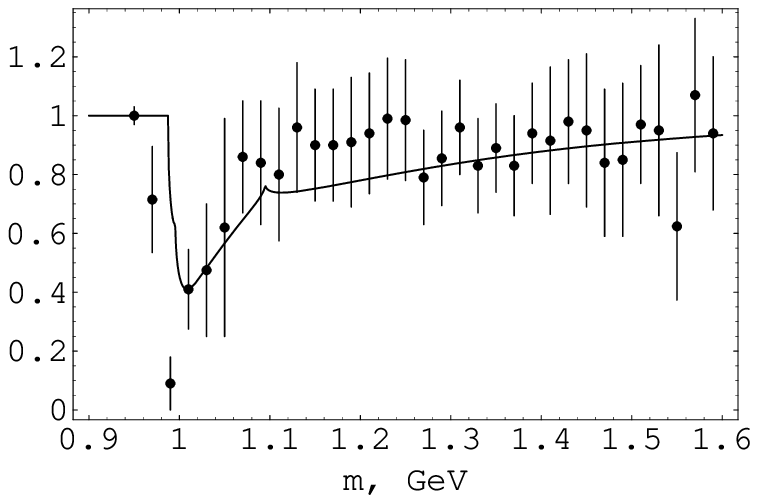}\\
\includegraphics[width=6pc]{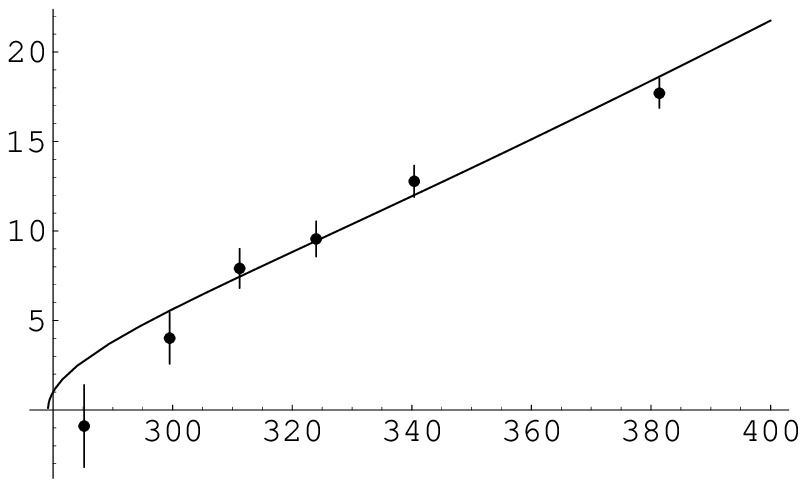}\ \
\includegraphics[width=6pc]{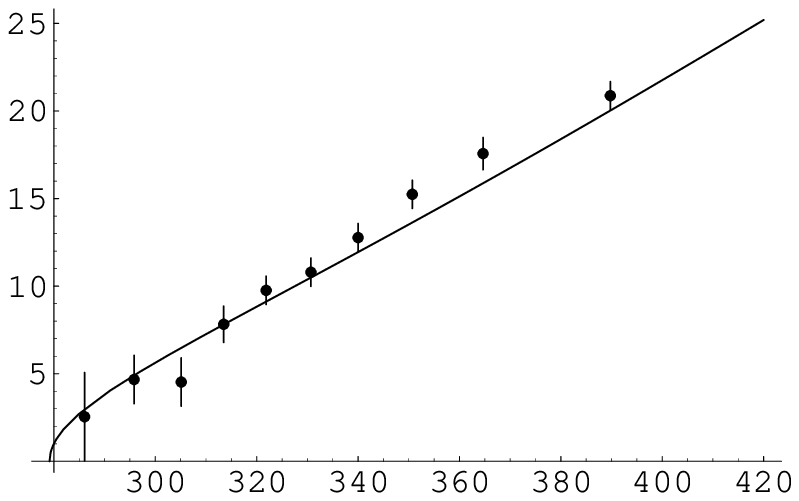}\\
\includegraphics[width=6pc]{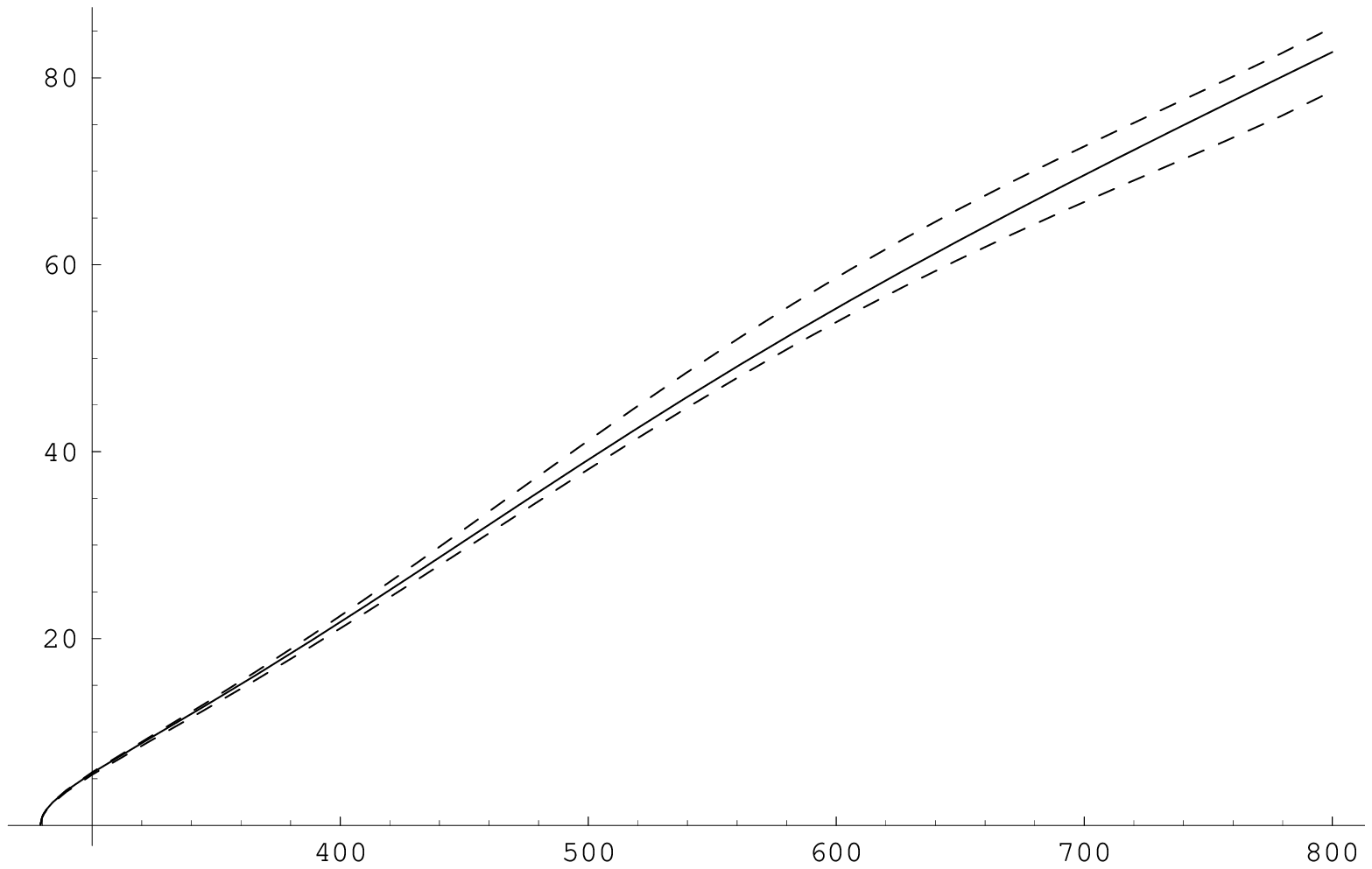}\ \
\includegraphics[width=6pc]{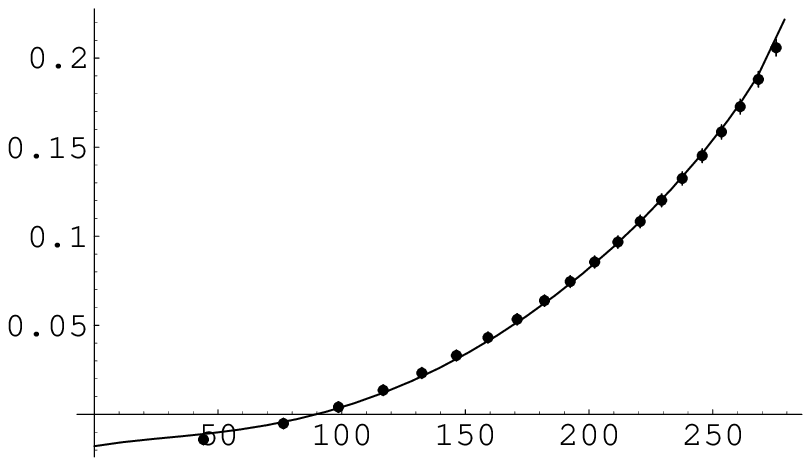}
\end{tabular}\\
\end{center}
{\footnotesize Figure 5. The phenomenological chiral shielding.
$\delta_0^0$=$\delta_B^{\pi\pi}$+$\delta_{res}$. The comparison with
the CERN-Munich data for $\delta^0_0$ and inelasticity $\eta^0_0$,
with the BNL and NA48 data for $\delta^0_0$, with the CGL band for
$\delta^0_0$ \cite{CGL}, and with Leutwyler's calculation of $T^0_0$
for $s<4m^2_\pi$, respectively.}
\end{figure}

\section{Four-quark model \cite{Ja,A1,A2}}
The nontrivial nature of the well-established light scalar
resonances $f_0(980)$ and $a_0(980)$ is no longer denied practically
anybody. In particular, there exist numerous evidences in favour of
the $q^2\bar q^2$ structure of these states. As for the nonet as a
whole, even a dope's look at PDG Review gives an idea of the
four-quark structure of the light scalar meson nonet,\,\footnote{To
be on the safe side, notice that LSM does not contradict to
non-$q\bar q$ nature of the low lying scalars because Quantum Fields
can contain different virtual particles in different regions of
virtuality.} $\sigma(600)$, $\kappa$(700-900), $a_0(980)$, and
$f_0(980)$, inverted in comparison with the classical $P$ wave
$q\bar q$ tensor meson nonet $a_2(1270)$, $f_2(1320)$,
$K_2^\ast(1420)$, and $f_2^\prime (1525)$. Really, while the scalar
nonet cannot be treated as the $P$ wave $q\bar q$ nonet in the naive
quark model, it can be easy understood as the $q^2\bar q^2$ nonet,
where $\sigma(600)$ has no strange quarks, $\kappa$(700-900) has the
$s$ quark, $a_0(980)$ and $f_0(980)$ have the $s\bar s$ pair. The
scalar mesons $a_0(980)$ and $f_0(980)$, discovered more than thirty
years ago, became the hard problem for the naive $q\bar q$ model
from the outset. Really, on the one hand, the almost exact
degeneration of the masses of the isovector $a_0(980)$ and isoscalar
$f_0(980)$ states revealed seemingly the structure similar to the
structure of the vector $\rho$ and $\omega$ or tensor $a_2(1320)$
and $f_2(1270)$ mesons, but on the other hand, the strong coupling
of $f_0(980)$ to $K\bar K$ suggests a considerable $s\bar s$ part in
the $f_0(980)$ wave function.

In 1977 Jaffe noted that in the MIT bag model, which incorporates
confinement phenomenologically, there are light four-quark scalar
states. He suggested also that $a_0(980)$ and $f_0(980)$ might be
these states. From that time $a_0(980)$ and $f_0(980)$ resonances
came into beloved children of the light quark spectroscopy.

\section{Radiative decays of \boldmath $\phi$ meson \cite{A1,A2,AI,AG97,AG01}}
Ten years later we showed \cite{AI} that the study of the radiative
decays $\phi\to\gamma a_0\to\gamma\pi\eta$ and $\phi\to\gamma f_0\to
\gamma\pi\pi$ can shed light on the problem of $a_0(980)$ and
$f_0(980)$ mesons. Over the next ten years before experiments (1998)
the question was considered from different points of view. Now these
decays have been studied not only theoretically but also
experimentally. The first measurements were reported by SND and
CMD-2. More recently KLOE performed measurements which are in close
agreement with the Novosibirsk data but have considerably smaller
errors. Note that $a_0(980)$ is produced in the radiative $\phi$
meson decay as intensively as $\eta '(958)$ containing $\approx 66\%
$ of $s\bar s$, responsible for $\phi\approx s\bar s\to\gamma s\bar
s\to\gamma \eta '(958)$. It is a clear qualitative argument for the
presence of the $s\bar s$ pair in the isovector $a_0(980)$ state,
i.e., for its four-quark nature.

\section{\boldmath
$K^+K^-$ loop mechanism of creation, spectra, and gauge invariance
\cite{AK06,A1,A2,AI,AG97,AG01,AK03}} When basing the experimental
investigations, we suggested \cite{AI} one-loop model $\phi\to
K^+K^-\to\gamma a_0(980)$ (or $f_0(980))$; see Fig. 6. This model is
used in the data treatment and is ratified by experiment. Below we
argue on gauge invariance grounds that the present data give the
conclusive arguments in favor of the $K^+K^-$ loop transition as the
principal mechanism of $a_0(980)$ and $f_0(980)$ production in the
$\phi$ radiative decays. This enables to conclude that production of
the lightest scalars $a_0(980)$ and $f_0(980)$ in these decays is
caused by the four-quark transitions, resulting in strong
restrictions on the large $N_C$ expansions of the decay amplitudes.
The analysis shows that these constraints give new evidences in
favor of the four-quark nature of $a_0(980)$ and $f_0(980)$ mesons.
The data are described in the model $\phi\to(\gamma
a_0+\pi^0\rho)\to\gamma\pi^0\eta$ and $\phi\to [\gamma
(f_0+\sigma)+\pi^0\rho]\to\gamma\pi^0\pi^0$; see Figs. 7.

To describe the experimental spectra $dBR(\phi\to\gamma R\to\gamma
ab,m)/dm$=$|g_R(m)|^2\omega (m) p_{ab}(m)|g_{Rab}/$\\
$D_R(m)|^2/(\Gamma_\phi\, 48\pi^3m_{\phi}^2)$, $R$=$a_0,f_0$,
$ab$=$\pi^0\eta, \pi^0\pi^0$, the function $|g_R(m)|^2$ should be
smooth (almost constant) for $m\leq 0.99$ GeV. But the problem
issues from gauge invariance which requires that $g_R(m)$ is
proportional to the photon energy
$\omega(m)$=$(m_{\phi}^2-m^2)/2m_{\phi}$ (at least!) in the soft
photon region. Stopping the function $(\omega (m))^2$ at $\omega
(990\,\mbox{MeV})$=29 MeV with the help of the form-factor
$1/[1+(R\omega (m))^2]$ requires a prohibitive $R\approx100$
GeV$^{-1}$. $R\approx10$ GeV$^{-1}$ allows us to obtain the maximum
of the mass spectrum only near 900 MeV. The $K^+K^-$ loop model
$\phi\to K^+K^-\to\gamma R$ solves this problem in the elegant way:
a fine threshold phenomenon is discovered, see Fig. 8.

\section{\boldmath $K^+K^-$ loop mechanism \cite{A2,AG01}}
In truth this means that $a_0(980)$
and $f_0(980)$ are seen in the radiative decays of $\phi$ meson
owing to $K^+K^-$ intermediate state. So, the mechanism of
production of $a_0(980)$ and $f_0(980)$ mesons in the $\phi$
radiative decays is established at a physical level of proof. The
real  part of the $\phi\to\gamma R$ amplitude contains two different
contribution. One is caused by intermediate momenta (a few GeV) in
the loops and the other is caused by super high momenta in the
loops. At $\omega(m)$=0 these contribution eliminate each other.
With increasing $\omega(m)$ the contribution from intermediate
momenta decreases rapidly enough. The contribution from super high
momenta is constant and causes the $\phi\to\gamma R$ amplitude in a
great part.

\begin{figure}\begin{center}
\begin{tabular}{ccc}
\includegraphics[width=4.3pc]{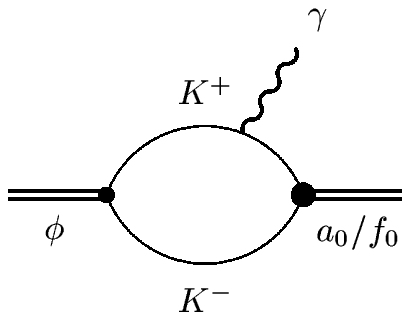}\
\raisebox{-3mm}{$\includegraphics[width=4.3pc]{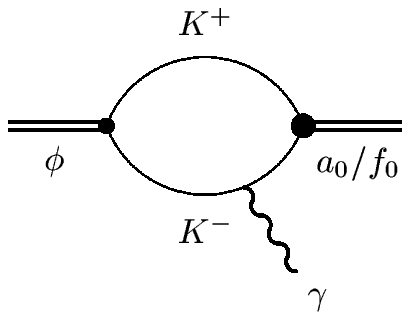}$}\
\includegraphics[width=4.3pc]{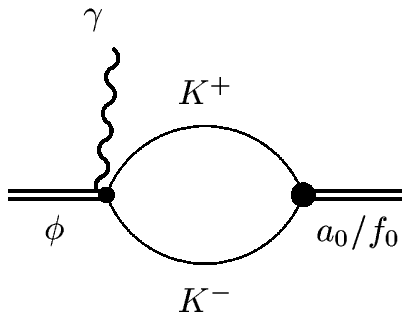}\\
{\footnotesize Figure 6. The $K^+K^-$ loop model.}
\end{tabular}
\end{center}
\begin{center}
\begin{tabular}{cc}
\includegraphics[width=8pc,height=5.6pc]{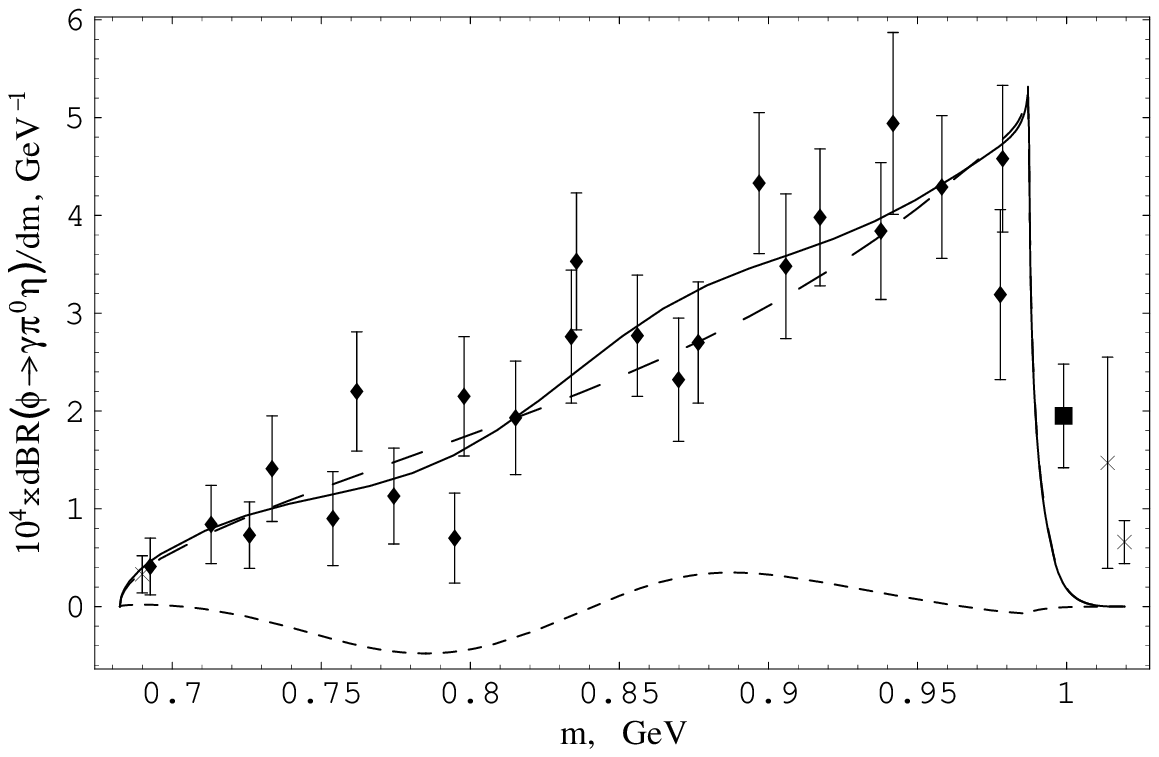}\
\includegraphics[width=8pc,height=5.6pc]{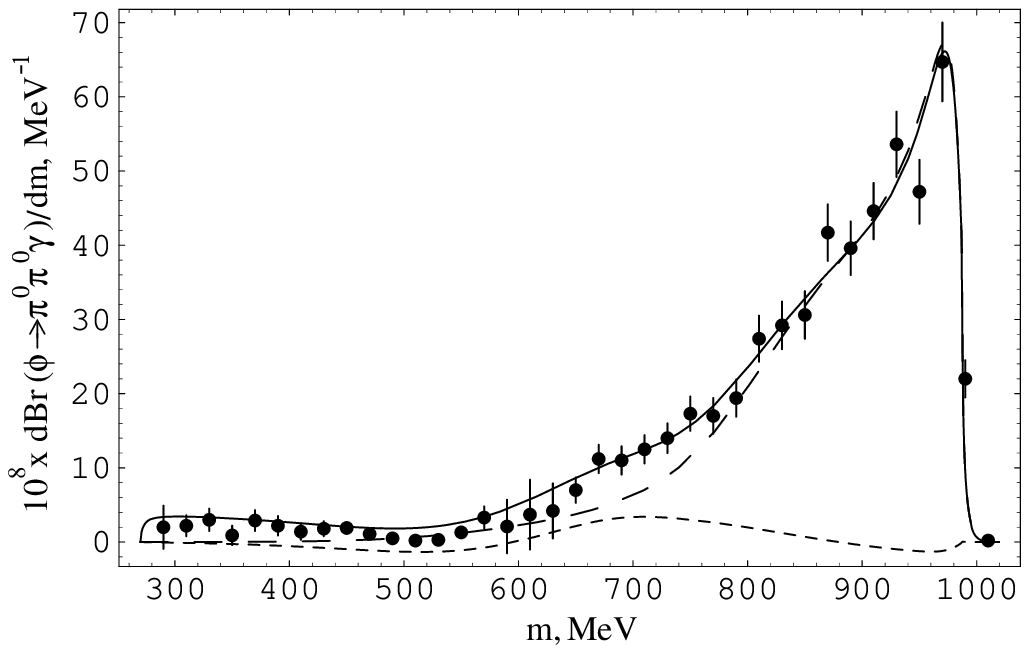}
\end{tabular}\\
\end{center}
{\footnotesize Figure 7. The left (right) plot illustrates the fit
to the KLOE data for the $\pi^0\eta$ ($\pi^0 \pi^0$) mass spectrum
in the $\phi\to\gamma\pi^0\eta$ ($\phi\to\gamma\pi^0\pi^0$) decay.}
\begin{center}
\begin{tabular}{c}
\includegraphics[height=5.6pc]{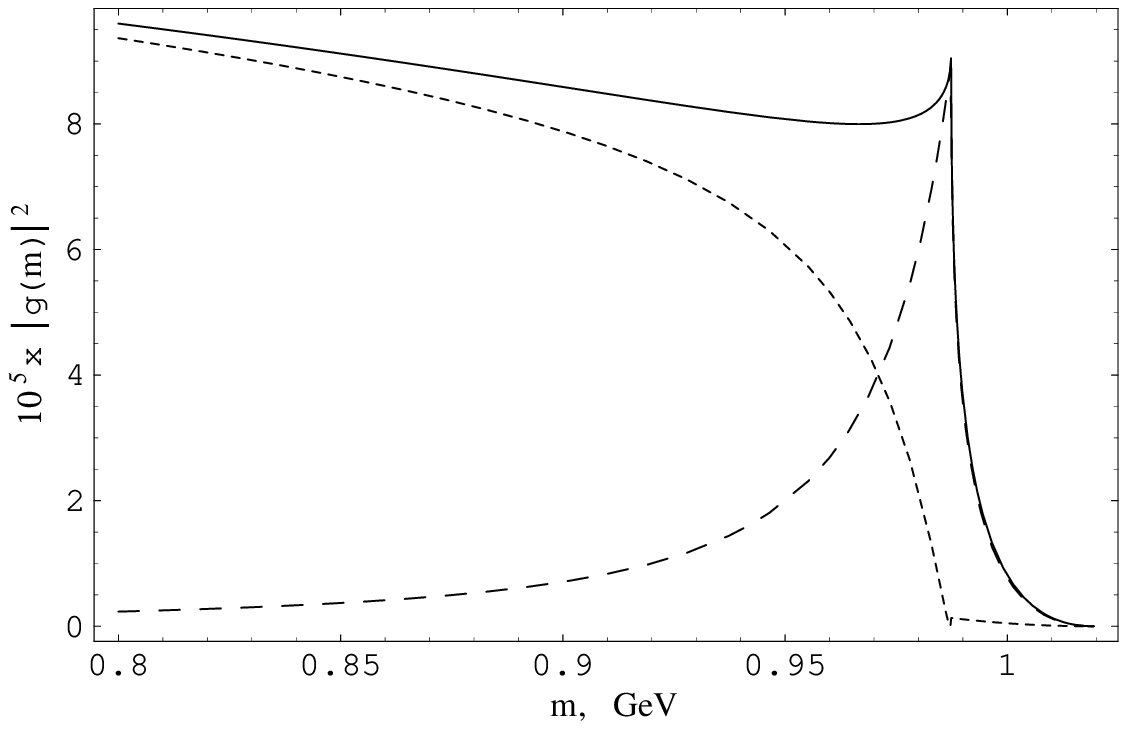}\end{tabular}\\
\end{center} {\footnotesize Figure 8. A new threshold phenomenon in
$\phi\to K^+K^-\to\gamma R$ decays. The universal in $K^+K^-$ loop
model function $|g(m)|^2= |g_R(m)/g_{RK^+K^-}|^2$ is drawn with the
solid line. The contributions of the imaginary and real parts of
$g(m)$ are drawn with the dashed and dotted lines,
respectively.}\end{figure}

\section{Four-quark transition and OZI \cite{A2}}
Both real and imaginary parts of the $\phi\to\gamma R$ amplitude are
caused by the $K^+K^-$ intermediate state. The imaginary part is
caused by the real $K^+K^-$ 
state while the real part
is caused by the virtual compact $K^+K^-$ 
state, i.e., we are dealing here with the four-quark transition.
Needless to say, radiative four-quark transitions can happen between
two $q\bar q$ states as well as between $q\bar q$ and $q^2\bar q^2$
states but their intensities depend strongly on a type of the
transitions. A radiative four-quark transition between two $q\bar q$
states requires creation and annihilation of an additional $q\bar q$
pair, i.e., such a transition is forbidden according to the
Okubo-Zweig-Iizuka (OZI) rule, while a radiative four-quark
transition between $q\bar q$ and $q^2\bar q^2$ states requires only
creation of an additional $q\bar q$ pair, i.e., such a transition is
allowed according to the OZI rule. The consideration of this problem
from the large $N_C$ expansion standpoint supports the suppression
of  a radiative four-quark transition between two $q\bar q$ states
in comparison with a radiative four-quark transition between $q\bar
q$ and $q^2\bar q^2$ states.

\section{\boldmath $a_0(980)/f_0(980)\to\gamma\gamma$  \& $q^2\bar
q^2$ model} Recall that twenty six years ago the suppression of
$a_0(980)/f_0(980)\to\gamma\gamma$  was predicted \cite{ADS1,ADS2}
based on the $q^2\bar q^2$ model,
$\Gamma(a_0(980)/f_0(980)\to\gamma\gamma)\sim 0.27\,\mbox{keV}$.
Experiment supported this prediction $\Gamma
(a_0\to\gamma\gamma)$=$(0.19\pm 0.07
^{+0.1}_{-0.07})/B(a_0\to\pi\eta)$ keV, Crystal Ball, and $(0.28\pm
0.04\pm 0.1)/B(a_0\to\pi\eta)$ keV, JADE, $\Gamma
(f_0\to\gamma\gamma)$=$(0.31\pm0.14\pm0.09)$ keV, Crystal Ball, and
$(0.24\pm0.06\pm0.15)$ keV, MARK II. When in the $q\bar q$ model it
was anticipated $\Gamma(a_0\to\gamma\gamma)$=(1.5-5.9)$\Gamma
(a_2\to\gamma\gamma)$=(1.5-5.9)$(1.04\pm 0.09)$ keV and
$\Gamma(f_0\to\gamma\gamma)$=(1.7-5.5)$\Gamma (f_2\to\gamma\gamma)$=
(1.7-5.5)$(2.8\pm 0.4)$ keV. The $a_0\to K^+K^-\to\gamma\gamma$
model \cite{AS88} describes adequately data and corresponds to the
four-quark transition $a_0\to q^2\bar q^2\to\gamma\gamma$,
$\langle\Gamma(a_0\to K^+K^-\to\gamma\gamma\rangle\approx 0.3$ keV.

\section{\boldmath $\gamma\gamma\to\pi\pi$
from Belle \cite{Belle,AS05,AS08}} Recently, we analyzed the new
high statistics Belle data on the reactions
$\gamma\gamma$\,$\to$\,$\pi^+\pi^-$ and
$\gamma\gamma$\,$\to$\,$\pi^0\pi^0$, and clarified the current
situation around the $\sigma(600)$, $f_0(980)$, and $f_2(1270)$
resonances in $\gamma\gamma$ collisions. The new Belle data are
shown in Fig. 9, together with our fitted curves. The
$f_0$\,$\to$\,$K^+K^-$\,$\to$\,$\gamma\gamma$ approximation yields
$\langle\Gamma_{f_0\to
K^+K^-\to\gamma\gamma}\rangle$\,$\approx$\,0.2\,keV. The direct
couplings $\sigma$\,$\to$\,$\gamma\gamma$ and
$f_0$\,$\to$\,$\gamma\gamma$ are small:
$\Gamma^{direct}_{\sigma\to\gamma\gamma}\ll 0.1\mbox{\,keV},\
\Gamma^{direct}_{f_0\to\gamma\gamma} \ll 0.1\mbox{\,keV}$.
\begin{figure}
\begin{center}
\begin{tabular}{cc}
\includegraphics[width=8.1pc,height=7pc]{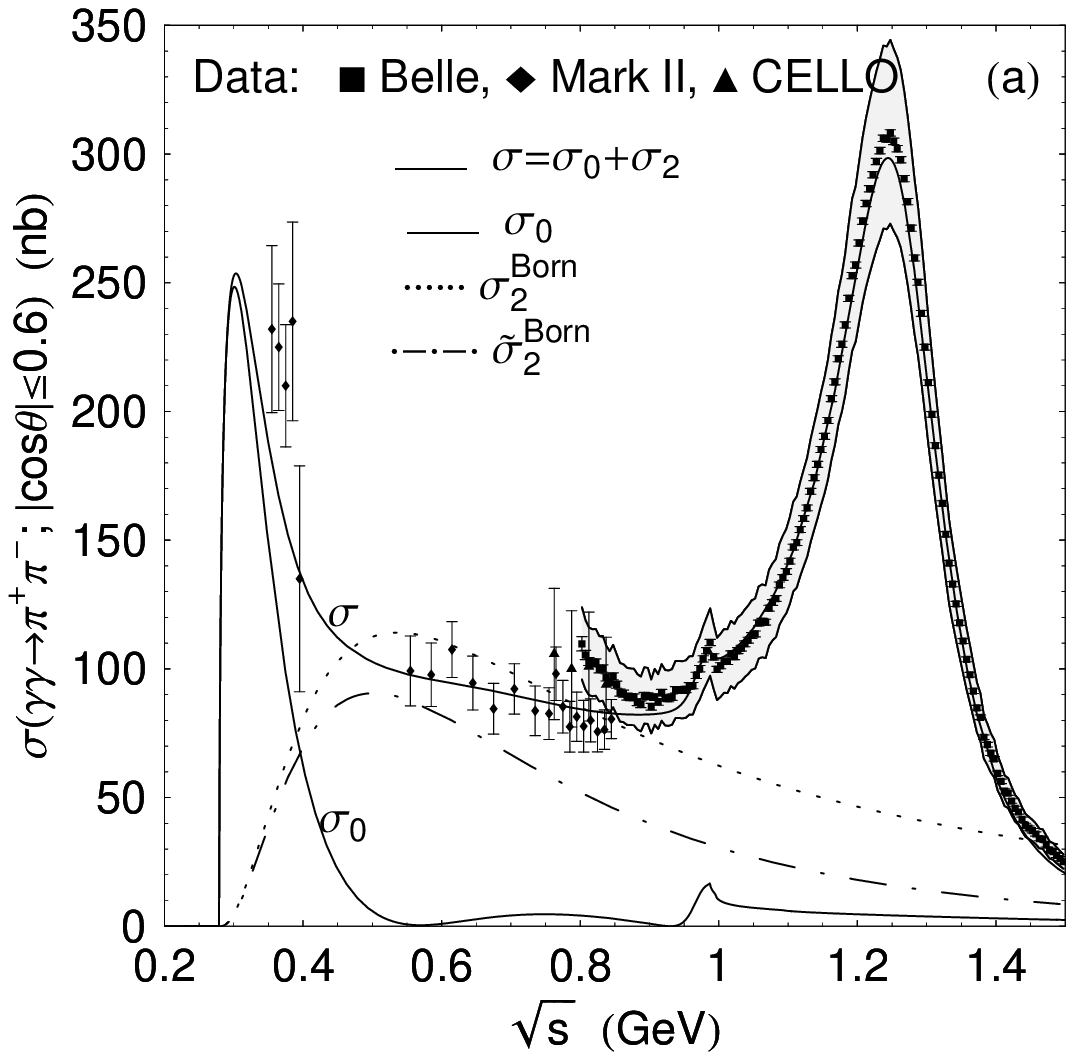}\ \
\includegraphics[width=8.1pc,height=7pc]{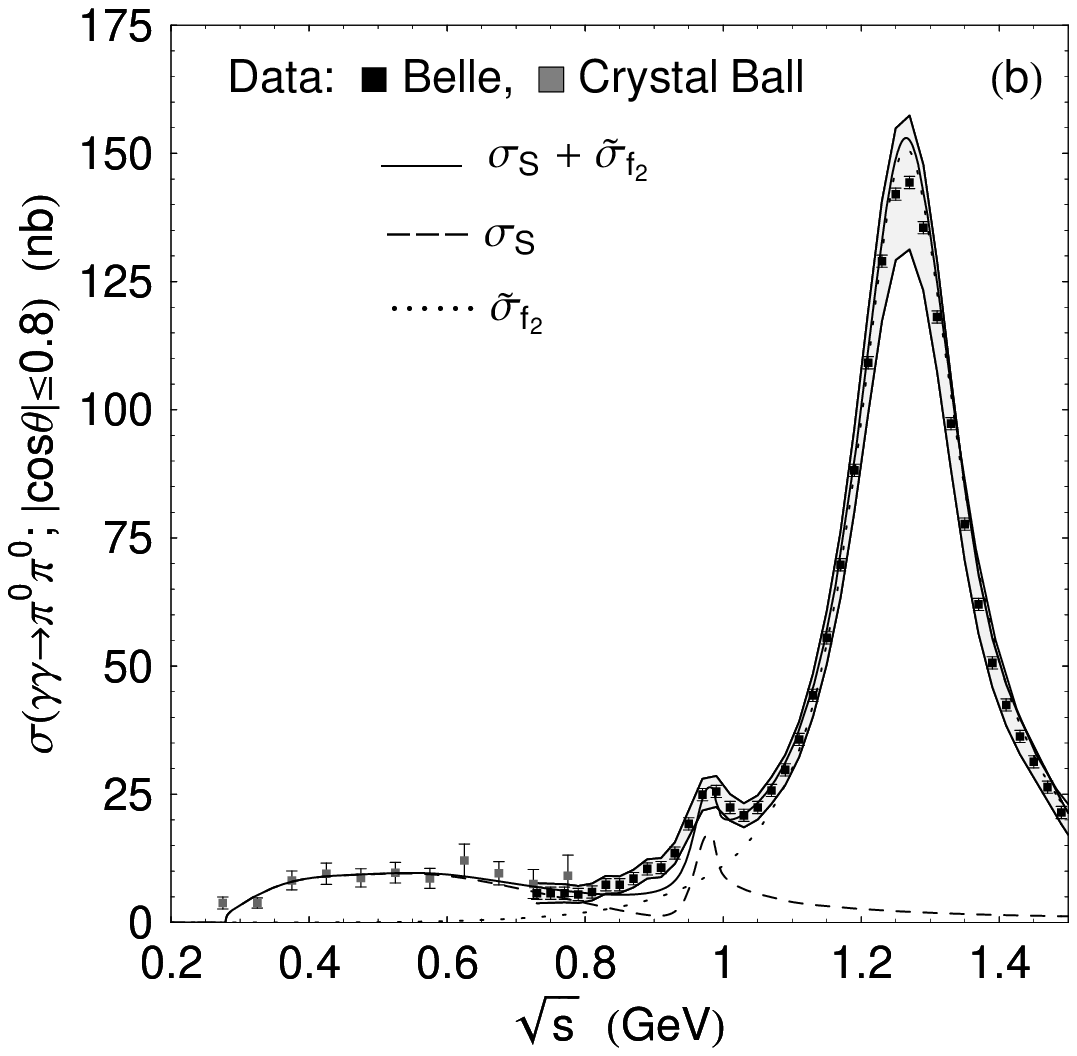}
\end{tabular}
\\ \end{center} {\footnotesize Figure 9. (a) Cross section for $\gamma\gamma\to\pi^+
\pi^-$. (b) Cross section for $\gamma\gamma\to\pi^0\pi^0$.}
\end{figure}

We thank Heiri Leutwyler very much for communications. This work was
supported in part by the Presidential Grant No. NSh-1027.2008.2 and
by the RFFI Grant No. 07-02-00093.

\end{document}